\documentclass[floatfix, prl, nopacs, twocolumn,showkeys, preprintnumbers, nofootinbib, superscriptaddress]{revtex4-1}
\usepackage[utf8]{inputenc}
\usepackage[sort&compress]{natbib}
\usepackage{ulem}
\usepackage{overpic}
\usepackage{bm}
\usepackage{times}
\usepackage{amssymb,amsbsy,amsmath,amsfonts}
\usepackage{graphicx}
\usepackage{float}
\usepackage{lipsum}
\usepackage{color}
\usepackage{booktabs}
\usepackage{makecell}
\usepackage{rotating}
\usepackage{srcltx}
\usepackage{slashed}
\usepackage{subfigure}
\usepackage{multirow}
\usepackage{verbatim}
\usepackage{hyperref}
\usepackage{tabularx}

\DeclareUnicodeCharacter{3002}{HEREHEREHERE}

\begin{document}

\title{Implication of the existence of   a $J^{PC}=0^{--}$ $\bar{D}_sDK$  bound state  on the nature of $D_{s0}^*(2317)$
and new configuration of exotic state }

\author{Tian-Wei Wu}
\affiliation{School of Science, Shenzhen Campus of Sun Yat-sen University, Shenzhen 518107, China}

\author{Ming-Zhu Liu}
\email[Corresponding author: ]{liumz@lzu.edu.cn}
\affiliation{
Frontiers Science Center for Rare isotopes, Lanzhou University,
Lanzhou 730000, China}
\affiliation{ School of Nuclear Science and Technology, Lanzhou University, Lanzhou 730000, China}

\author{Li-Sheng Geng}\email[Corresponding author: ]{lisheng.geng@buaa.edu.cn}
\affiliation{School of Physics,  Beihang University, Beijing 102206, China}
\affiliation{Sino-French Carbon Neutrality Research Center, \'Ecole Centrale de P\'ekin/School of General Engineering, Beihang University, Beijing 100191, China}
\affiliation{Peng Huanwu Collaborative Center for Research and Education, Beihang University, Beijing 100191, China}
\affiliation{Beijing Key Laboratory of Advanced Nuclear Materials and Physics, Beihang University, Beijing 102206, China }
\affiliation{Southern Center for Nuclear-Science Theory (SCNT), Institute of Modern Physics, Chinese Academy of Sciences, Huizhou 516000, Guangdong Province, China}

\begin{abstract}
The discovery of numerous new hadrons over the past two decades has provided unprecedented opportunities to understand the non-perturbative QCD and hadron structure.
Hadronic molecule picture plays an important role in explaining these new hadrons and enriching the configurations of exotic hadronic states. 
In this letter, using the model-independent $DK$ potential extracted from the relevant experimental data, a $J^{PC}=0^{--}$ $\bar{D}_sDK$ three-body hadronic molecule is predicted with a mass of $4310^{+14}_{-24}$ MeV. This state   shows decoupling to conventional   $c\bar{c}$ charmonia or the $\bar{D}_s D_{s0}^*(2317)$ two-body molecular state. It can be regarded  as a compelling three-body hadronic molecular candidate.    We further demonstrate that the $B^+ \to {D}^{*\pm}D^\mp K^+$  decays could be promising channels for searching for the predicted  state in future high-luminosity LHCb runs.  

\end{abstract}
\date{\today}


\maketitle

\noindent{\it Introduction.—}
Quantum chromodynamics (QCD), the fundamental theory of the strong interaction, exhibits strong coupling at low energies, resulting in color confinement. In this regime, the relevant degrees of freedom are hadrons rather than quarks and gluons, making quark-level studies of low-energy strong interactions particularly challenging.  Consequently, hadron spectroscopy has become indispensable for investigating non-perturbative strong interactions, especially in light of the numerous new heavy hadrons discovered since $2003$.  
A remarkable feature of the spectrum of these heavy hadrons is that most are near the threshold of a pair of hadrons. 
The recent studies from the unquenched quark model~\cite{Ortega:2016mms,Albaladejo:2018mhb,Luo:2021dvj,Yang:2021tvc},  effective field theories~\cite{Cheng:2014bca,MartinezTorres:2014kpc,Guo:2015dha,Yao:2015qia,Guo:2023wkv,Gil-Dominguez:2023huq}, and lattice QCD~\cite{Liu:2012zya,Mohler:2013rwa,Lang:2014yfa,Bali:2017pdv,Alexandrou:2019tmk} indicate that the heavy hadrons have strong couplings to a pair of hadrons, where the hadron-hadron interactions represent a fundamental manifestation of non-perturbative QCD dynamics.  Importantly, such non-perturbative effects are not limited to spectroscopy but also emerge prominently in both the final-state interactions of heavy hadron decays~\cite{Guo:2018kno,Fu:2021wde} and their production~\cite {Navarra:2015iea,Albaladejo:2016hae,Liu:2022dmm}. Therefore, the hadron-hadron interactions are crucial for understanding the properties of heavy hadrons.    

A salient example is the $DK$ interaction, which has garnered renewed interest following the discovery of the $D_{s0}^*(2317)$~\cite{BaBar:2003oey}. If   $D_{s0}^*(2317)$ is interpreted as a    $c\bar{s}$ charmed strange meson with  $J^P=0^+$,  its mass is  around  $160$  MeV    lower than the theoretical predictions from the Godfrey-Isgur model~\cite{Godfrey:1985xj}. The mass puzzle of  $D_{s0}^*(2317)$ is solved if the $DK$ molecular component is embodied~\cite{Albaladejo:2018mhb,Yang:2021tvc,Luo:2021dvj,Hao:2022vwt,Yang:2023tvc,Ni:2023lvx}. Lattice QCD calculations of the $DK$ scattering parameters~\cite{Liu:2012zya,Bali:2017pdv,Cheung:2020mql} reveal that the molecular component constitutes over  $70\%$ of the  $D_{s0}^*(2317)$ wave function~\cite{MartinezTorres:2014kpc,Albaladejo:2018mhb,Yang:2021tvc,Guo:2023wkv,Gil-Dominguez:2023huq}. However, the limited experimental data on  $D_{s0}^*(2317)$ has hindered independent verification of this picture through alternative observables. Recent theoretical studies have proposed investigating the $DK$ interaction and  $D_{s0}^*(2317)$ properties within three-body hadron systems~\cite{Liu:2024uxn}, inspiring many studies on analogous three-body hadronic molecules~\cite{MartinezTorres:2009xb,Ma:2017ery,Wu:2019vsy,MartinezTorres:2020hus,Wei:2022jgc,Tan:2024omp,Zhang:2024yfj,Liu:2024uxn}. These three-body hadronic molecules represent novel configurations of hadronic matter that offer unique insights into non-perturbative QCD dynamics. These studies not only deepen our understanding of hadron structure but may also establish new paradigms in hadron spectroscopy.

Originally,  the Valencia group employed the  Fixed Center Approximation to the Faddeev equations and predicted the existence of two exotic hadrons: three-body  $NDK$~\cite{Xiao:2011rc}  and $DK\bar{K}$~\cite{Debastiani:2017vhv} molecules, where the $DK$ interaction is determined by reproducing the $D_{s0}^*(2317)$ mass. However, verifying the $DK$ interaction via the above three-body hadron systems is less optimal since the $DN$ and $K\bar{K}$ potentials are strongly attractive~\cite{Xiao:2011rc,Debastiani:2017vhv}.  This motivated our investigation of the  $DDK$ system,  where the weak $DD$ interaction allows the $DK$ component to dominate the molecular binding~\cite{MartinezTorres:2018zbl,Wu:2019vsy,Pang:2020pkl}.  Although theoretically promising, the $DDK$ molecule presents experimental challenges: its production in $e^+e^-$ collisions is highly suppressed~\cite{Belle:2020xca}, although it is possible in $B_c$ decays~\cite{Liu:2024uxn}. However, this pathway suffers from the intrinsically low $B_c$ production cross section.  Subsequently, a $D\bar{D}K$ molecule was also predicted~\cite{Wu:2020job}, but its yield in the inclusive process of $e^+e^-$ collisions is lower by three orders of magnitude than that of the two-body $DK$ molecule~\cite {Wu:2022wgn}. In addition, other components can mix with the three-body hadronic molecule~\cite{SanchezSanchez:2017xtl}. The current landscape thus lacks an unambiguous three-body molecular candidate that simultaneously satisfies two crucial criteria:  minimal contamination from other hadronic configurations and experimentally accessible production rates. The discovery of such a state would provide decisive evidence for three-body hadronic molecules and offer unique insights into the fundamental interactions between hadrons.    

In this letter, we predict   a $J^{PC}=0^{--}\bar{D}_sDK$   three-body molecule with a binding
energy of about a few tens of MeV.     Such a quantum number is forbidden for conventional $c\bar{c}$ charmonia, ensuring negligible mixing with ordinary quarkonium states~\cite{Ji:2022blw}. Notably, while a $J^{PC}=0^{--}$ ${D}_{s0}^*\bar{D}_s$ two-body configuration is unlikely to form a bound state~\cite{Karliner:2016ith,Shen:2010ky}, the three-body  $\bar{D}_sDK$ system exhibits binding, further distinguishing it from possible two-body molecular interpretations.    Remarkably, this exotic state can be produced through $B$ decays.  The observation in the  decays  $B^+ \to {D}^{*\pm}D^{\mp}  K^+ $  by the LHCb Collaboration~\cite{LHCb:2024vfz} are promising, considering its decay behavior and production rates.    
In the following, we will first demonstrate the existence of the $J^{PC}=0^{--}$ $\bar{D}_sDK$ bound state with its interaction inputs, denoted by $X(4310)$.    Subsequently, we systematically address its decay properties and production mechanisms in $B$ decays.

 \begin{figure}[!h]
	\centering
	\includegraphics[width=8.0cm]{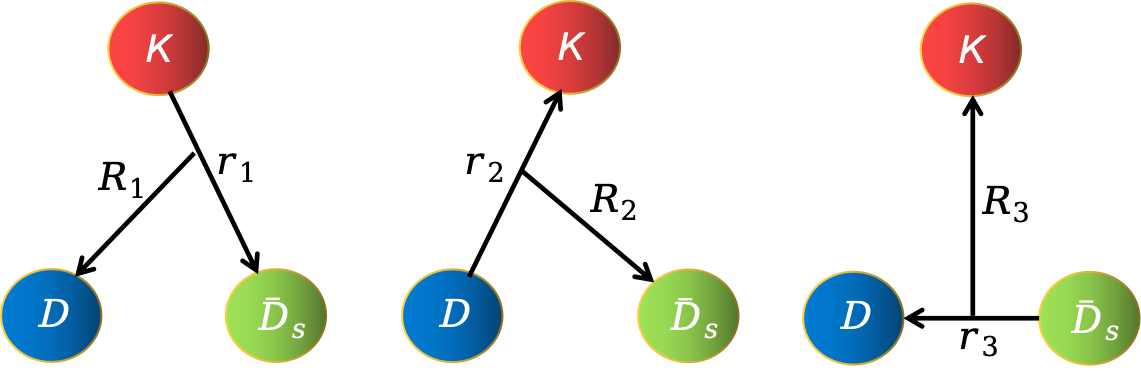}
	\caption{Three permutations of the Jacobi coordinates for the $\bar{D}_sDK$ system.   }\label{JAC}
\end{figure}

\noindent{\it Existence of a $J^{PC}=0^{--}$ $\bar{D}_sDK$ bound state $X(4310)$.—}
The three-body wave function for the $\bar{D}_sDK$ system of good $C$ parity reads
\begin{equation}\label{schd}
\Psi^{C}=\frac{1}{\sqrt{2}}(\Psi_{\bar{D}_sDK}+C\Psi_{D_s\bar{D}\bar{K}}'),
\end{equation}
where  $C=\pm1$ represents the $C$-parity eigenvalue,  and $\Psi$ ($\Psi'$) denotes the wave function of the $\bar{D}_{s}DK$ (${D}_{s}\bar{D}\bar{K}$) system.
The wave function $\Psi^C$ can be solved by the Schr\"odinger equation with the Hamiltonian $H=T+T'+V+V'+V^C$,  where $T$ ($T'$) and $V$ ($V'$) are the kinetic energy term and the hadron-hadron potentials of $\Psi$ ($\Psi'$), respectively.

\begin{figure}[ttt]
	\centering
	\includegraphics[width=7.5cm]{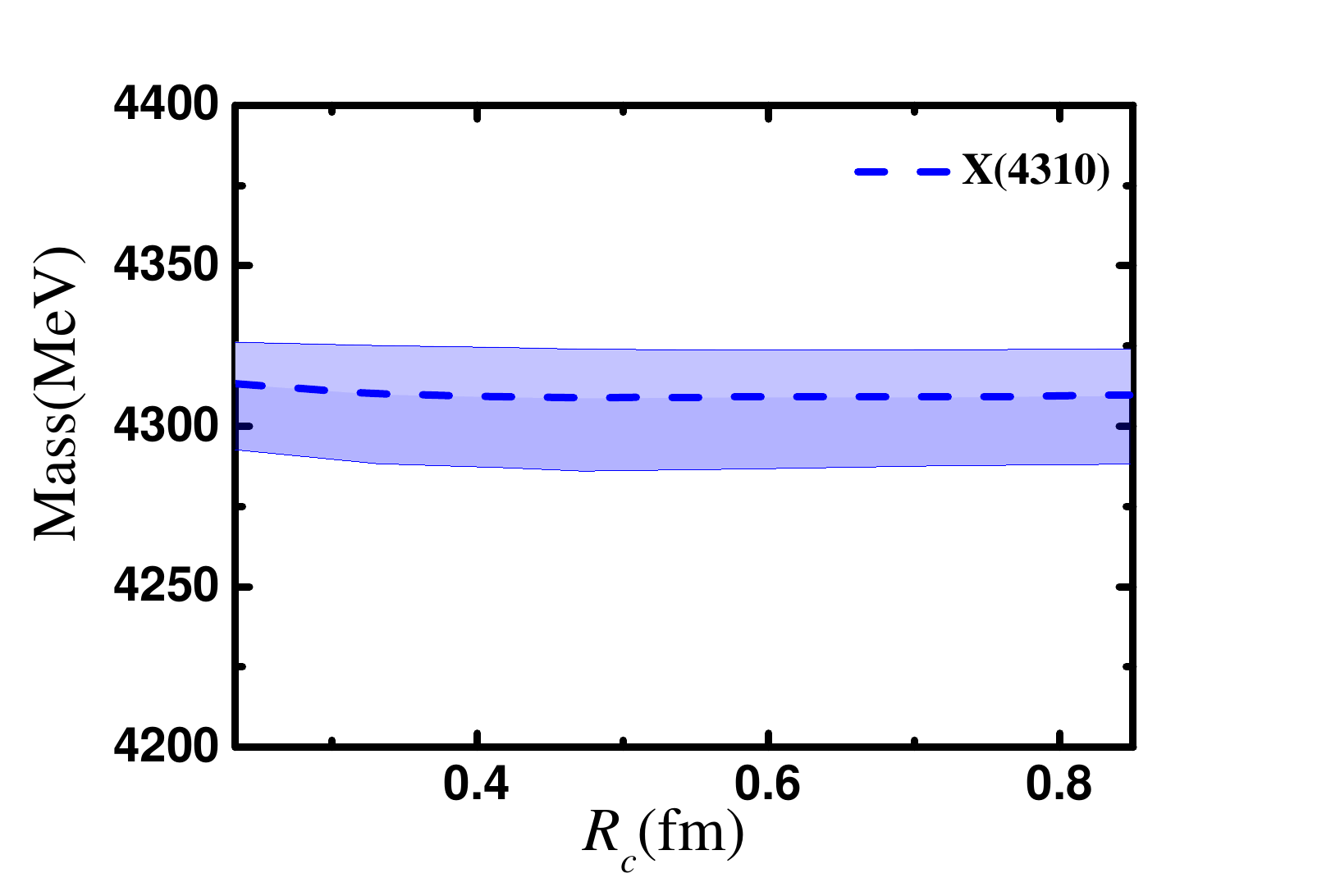}
	\caption{Mass of $X(4310)$ as a function of the cutoff $R_c$.   }\label{B.E.vsR_c}
\end{figure}

 In this letter, we employ the contact-range effective field theory (EFT) to construct the $DK$, $\bar{D}_sK$, and $D\bar{D}_s$ potentials~\cite{Hidalgo-Duque:2012rqv,Liu:2023cwk}. 
The contact-range potential in momentum space can be parameterized by a Gaussian-shaped potential in coordinate space

\begin{equation}
\label{TBpoten}
    V(r)=C
    _a\frac{e^{(r/R_c)2}}{\pi^{3/2}R_c^3},
\end{equation}
 where $R_c$ is a cut-off radius of the order of
a typical hadronic size.
 From systematic studies of exotic state masses and applications of SU(3)-flavor symmetry relations~\cite{Guo:2006fu,Guo:2009ct,Liu:2012zya,Altenbuchinger:2013vwa,Ji:2022vdj}, we establish the following approximate ratio for the contact potentials for a cutoff $\Lambda=1$~GeV: $C_a^{DK}:C_a^{\bar{D}_s K}: C_a^{\bar{D}_s D}=1:0.5:0.1$. Given the dominant role of the $DK$ interaction,  we focus particularly on determining this potential by precisely reproducing the $D_{s0}^*(2317)$ mass. 

We assume the   $D_{s0}^*(2317)$  as a mixture of a $DK-D_s\eta$ molecular state and a $c\bar{s}$ bare state rather than a pure $DK$ molecule.   
The lattice QCD simulations or reanalyses of lattice QCD results~\cite{Liu:2012zya,Bali:2017pdv,Cheung:2020mql,MartinezTorres:2014kpc,Albaladejo:2018mhb,Yang:2021tvc,Guo:2023wkv,Gil-Dominguez:2023huq} found that the molecular component accounts for more than $70\%$ of the $D_{s0}^*(2317)$ wave function.  Therefore, we assume that the molecular and bare components account for $70\%$ and $30\%$ of the physical $D_{s0}^*(2317)$, respectively.  In the mixture picture, the extracted  $DK$ potential forms a bound state with a binding energy of $14$ MeV, which is less attractive than assuming the $D_{s0}^*(2317)$ as a pure $DK$ molecule.
To estimate the uncertainty of the extracted $DK$ potential, we vary the molecular compositeness of the $D_{s0}^*(2317)$ from $50\%$ to $100\%$.

 \begin{figure*}[ht]
	\centering
	\includegraphics[width=15cm]{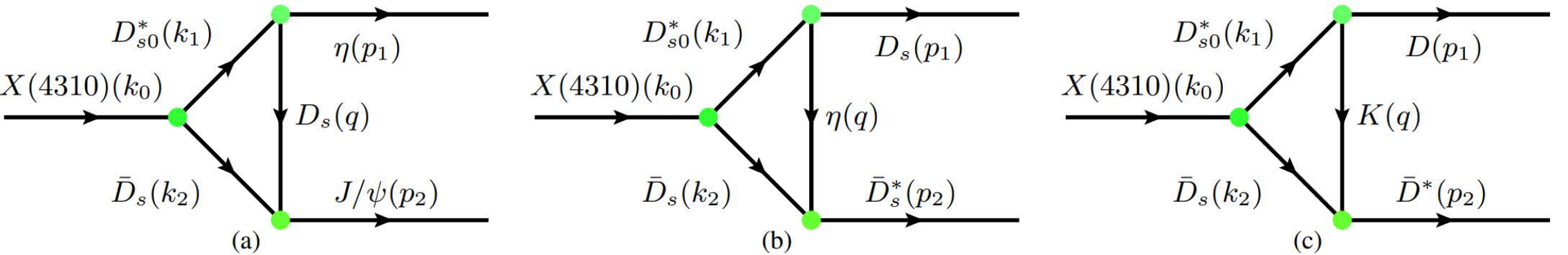}
	\caption{Triangle diagrams of the  $X(4310)$ via the subsystem $D_{s0}^*\bar{D}_s$  decaying into $J/\psi \eta$, $D_s\bar{D}_s^*$, and $D\bar{D}^*$.     }\label{triangle}
\end{figure*}

The potential $V^{C}$, dependent on the $C$-parity, could be a three-body interaction correlating the wave functions $\Psi$ and $\Psi'$. Here, since the $DK$ potential can form a bound state ${D}_{s0}^*$, we use an effective two-body $\bar{D}_sD_{s0}^*$ potential utilizing the one-boson-exchange (OBE) model~\cite{Liu:2010hf,Shen:2010ky}.  To account for the composite nature of the interacting hadrons, the dipole form factor is  usually introduced in the OBE model~\cite{Tornqvist:1993ng}:
\begin{eqnarray}
F(q,\Lambda,m_E)=(\frac{\Lambda^2-m_E^2}{\Lambda^2-q^2}).\label{formfactor}
\end{eqnarray}
 $m_E$ represents the mass of the exchanged particle, and $\Lambda$ is a cutoff parameter, which can be further parameterized as $\Lambda=\alpha\Lambda_{QCD}+ m_{E}$~\cite{Cheng:2004ru}. Following the studies of the heavy hadron decays~\cite{Cheng:2004ru,Yu:2017zst} and  the spectrum of hadronic molecular states~\cite{Liu:2024uxn},   we vary the parameter $\alpha$ from $1$ to $2$. The subsequent final-state interactions in the strong decays of the $X(4310)$ and  the production of the $X(4310)$ in $B$ decays  are described by the OBE model as well.  
The $C$-parity-dependent potential of $\bar{D}_s D_{s0}^*$ in the OBE model arises from the $\eta$-meson exchange, which is attractive for the $C=+$ configuration and repulsive for the $C=-$ configuration.  The $\eta$ exchange potential for the $\bar{D}_sD_{s0}^*$ system is written as 
\begin{small}
\begin{eqnarray}
V^{C=\pm}&=& \mp\frac{2}{3}\frac{k^2}{f_\pi^2}q_0^2(\frac{e^{-mr}-e^{-\Lambda r}}{4\pi r}-\frac{\Lambda^2-m^2}{8 \pi \Lambda}e^{-\Lambda r}),
\end{eqnarray}
\end{small}
where $q_0=m_{D_{s0}^*}-m_{D_s}$, $k=0.56$ and $f_\pi=130$~MeV~\cite{Liu:2010hf}.

\begin{table}[!h]
 \centering
 \caption{  Binding energy (in units of MeV), weights of Jacobi channels, root mean square radii (in units of fm), and expectation values of the Hamiltonian (in units of MeV) of the $0^{--}$ $\bar{D}_sDK$ molecule. 
 \label{threebody0--} }
 \begin{tabular}{c|  ccc c}
 \hline\hline
 Scenarios&  B.E.($0^{--})$ &    $ P_{\bar{D}_sK-D}$& $P_{DK-\bar{D}_s}$ & $ P_{\bar{D}_sD-K}$\\
 \hline
$\alpha=1$ & $22^{+23}_{-14}$& $11^{+1}_{-1}$ \%&$78^{-1}_{+2}$ \%&$11^{+0}_{-1}$ \%\\
 $\alpha=2$&  $20^{+22}_{-13}$& $10^{+1}_{-1} \%$& $80^{-1}_{+2}$ \%&$10^{+0}_{-1}$ \%\\
 \hline
  Scenarios&  $r_{\bar{D}_sK}$ &  $r_{DK}$  & $r_{\bar{D}_sD}$ &$\langle T \rangle$\\
 \hline
$\alpha=1$ & $1.6_{+0.9}^{-0.4}$& $1.2_{+0.5}^{-0.3}$  &$1.4_{+0.8}^{-0.3}$&$177_{-74}^{+81} $  \\
 $\alpha=2$& $1.7_{+1.4}^{-0.4}$& $1.2_{+0.7}^{-0.3}$   &$1.6_{+1.3}^{-0.5}$&$169_{-78}^{+82}$ \\
 \hline
  Scenarios& $ \langle V_{D_s\bar{K}}\rangle$~~~ &  $\langle V_{DK}\rangle~~~$  &   $ \langle V_{D_s\bar{D}}\rangle$&$ \langle V^{C=-}_{D_s\bar{D}^*_{s0}}\rangle$\\
 \hline
$\alpha=1$ & $-40_{+20}^{-25}$& $-147_{+62}^{-73}$&$-14_{+6}^{-7}$&$2_{-1}^{+0}$\\
 $\alpha=2$&  $-37_{+22}^{-25}$& $-143_{+64}^{-73}$&$-13_{+7}^{-6}$&$3_{-1}^{+2}$\\
 \hline
 \hline
 \end{tabular}
 \end{table}

With the pair-wise two-body interactions, the Schr\"odinger equation of $\Psi^C$ can be simplified as
\begin{equation}\label{schd1}
\langle \Psi|(T+V_{DK}+V_{\bar{D}_sK}+V_{\bar{D}_sD}+V^C-E)|\Psi\rangle=0,
\end{equation}
which can be solved by the Gaussian Expansion Method (GEM) ~\cite{Hiyama:2003cu}, with the three Jacobi channels as shown in Fig.~\ref{JAC}.

 The results are shown in Table~\ref{threebody0--}, the $0^{--}$ $\bar{D}_sDK$ system is predicted to be a bound state with a binding energy of $21^{+24}_{-14}$ MeV. Our results indicate that when the $DK-D_s\eta$ molecular component  of the $D_{s0}^*(2317)$ ranges from  $50\%$ to  $100\%$, the $\bar{D}_sDK$ system always remains bound.  The weight of each Jacobi channel $\Phi_i$  is calculated by $P_i=\langle\Psi|\hat{P}_i^G|\Psi\rangle$, with $|\Psi\rangle=\sum_i |\Phi_i\rangle$ and the generalized projection operator
    $\hat{P}_i^G=\sum_j |\Phi_i\rangle S^{-1}_{ij}\langle\Phi_j|$ suited for a non-orthogonal basis~\cite{PhysRevB.90.075128}, 
where $S_{ij}^{-1}$ is the element of the inverse of the overlap $S$-matrix of Jacobi channels with $S_{ij}=\langle\Phi_i|\Phi_j\rangle$ ($i,j=1-3$). Noting that the calculated weights $P_i$ should not be viewed as percentage contributions of individual components~\cite{RingSchuck1980}, but rather as the relative importance of the Jacobi channels in forming the three-body bound state.
Interestingly, the weights of Jacobi channels $i=1-3$ in Fig.~\ref{JAC} are stable, which are about 10\%, 80\%, and 10\% of $\bar{D}_sK-D$, $DK-\bar{D}_s$, and $\bar{D}_sD-K$,  respectively, varying  only $1\sim 2$ percent. 
This indicates the  $0^{--}$ three-body bound state is mainly contributed by the $(DK)-\bar{D}_s$ channel, weakly dependent on the molecular component of the $D_{s0}^*(2317)$. To test the impact of the cutoff on our results, the $X$ mass as a function of  the cutoff $R_c$  is shown in Fig.~\ref{B.E.vsR_c}, which indicates that its mass is weakly dependent on the cutoff. 
  The root mean square radii and expectation values of the Hamiltonian  of the $0^{--}$ $\bar{D}_sDK$ bound state are also presented in Table~\ref{threebody0--}, thereby providing a more intuitive demonstration of the spatial extent of the predicted molecular state and the relative contributions of individual interaction terms in the Hamiltonian to its formation.

The $J^{PC}=0^{-+}$ $\bar{D}_s DK$ three-body system is also predicted to be bound with a binding energy of $27^{+24}_{-17}$ MeV, which is slightly more bound than its $C$-parity counterpart $X(4310)$. 
The predicted $J^{PC}=0^{-+}$ $\bar{D}_s DK$ molecule can couple to two other configurations: an excited $0^{-+}$ $c\bar{c}$ state~\cite{Godfrey:1985xj} and a $0^{-+}$ $\bar{D}_s D_{s0}^*(2317)$  two-body molecule~\cite{Karliner:2016ith}.  
The intricate non-perturbative dynamics involving all three configurations require a more systematic study on the $J^{PC}=0^{-+}$ $\bar{D}_s DK$ molecule. 
In contrast, the  $X(4310)$ molecule exhibits no coupling to either conventional $c\bar{c}$ states nor the $\bar{D}_s D_{s0}^*(2317)$ two-body configuration. It can be regarded as a compelling 
three-body hadronic molecular candidate. Therefore, we mainly focus on the $X(4310)$ in this letter.

 \begin{figure}[!h]
	\centering
	\includegraphics[width=8.5cm]{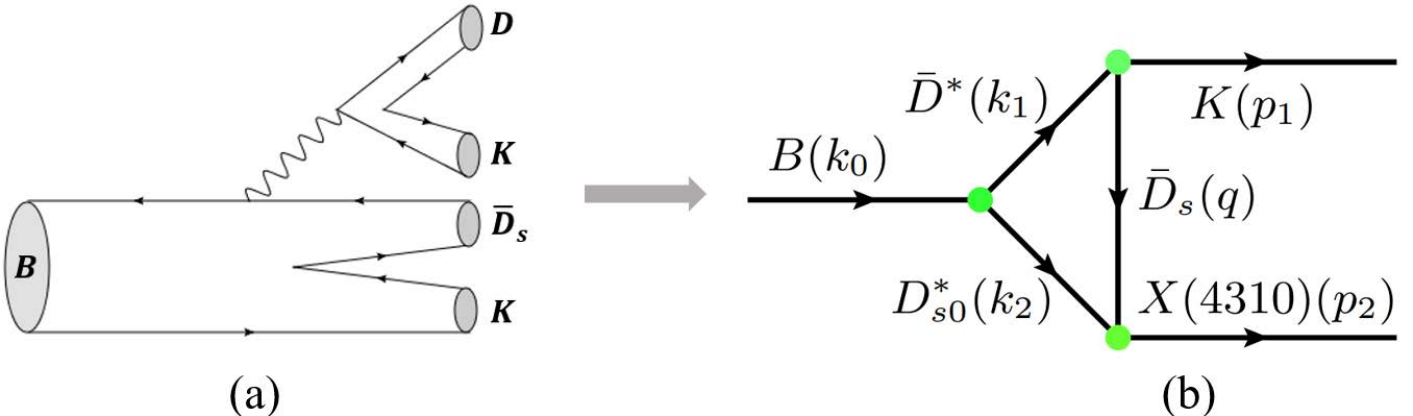}
	\caption{Topology diagram at the quark level (a)  and triangle diagram at the hadron level (b) accounting for the weak decays of $B \to X(4310) K$.    }\label{triangleproduction}
\end{figure}

\begin{figure*}[ht]
\centering
	\includegraphics[width=15cm]{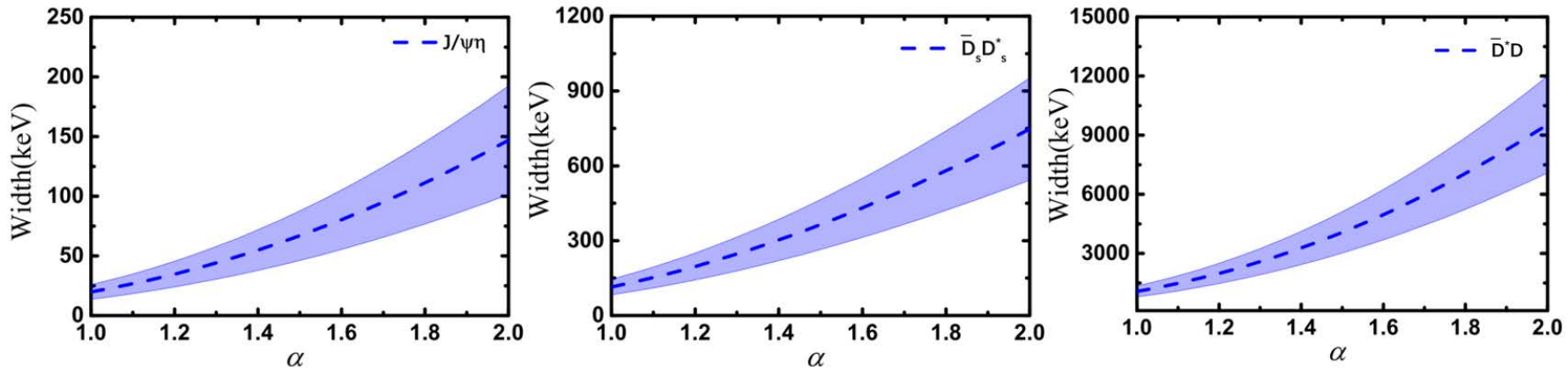}
	\caption{Partial decay widths  of $X(4310)\to J/\psi \eta$, $X(4310)\to \bar{D}_sD_s^*$, and $X(4310)\to \bar{D}^*D$ as functions of $\alpha$. The dashed line and band correspond to their central value and uncertainties.      }\label{decaywidths}
\end{figure*}

\noindent{\it Decays and productions of $X(4310)$—}      
 To facilitate potential observation of the predicted state $X(4310)$, we investigate its decay properties and production mechanisms through triangle diagrams, focusing particularly on $B$ decays~\cite{Cheng:2004ru,Faessler:2007gv}.  The Jacobi coordinate analysis (Fig.~\ref{JAC} and Table~\ref{threebody0--}) indicates that the $X(4310)$ predominantly decays via its  subsystem $\bar{D}_sD_{s0}^*$, which subsequently undergoes inelastic scattering into final states including   $J/\psi \eta$, $\bar{D}^*D$, and $\bar{D}_s^*D_s$ through meson-exchange processes (Fig.~\ref{triangle}). 
With the  Lagrangian for each vertex in Fig.~\ref{triangle}, we obtain the corresponding amplitudes as 
\begin{widetext}
\begin{eqnarray}
\label{amp1mae}
i\mathcal{M}_a&=&g_{XD_{s0}^*\bar{D}_{s}}g_{D_{s0}^*D_{s}\eta}g_{\psi \bar{D}_{s}D_{s}} \int\frac{d^{4}q}{(2\pi)^{4}}(k_{2}^{\mu}-q^{\mu})\frac{1}{k_{1}^{2}-m_{D_{s0}^*}^2}\frac{1}{k_{2}^{2}-m_{\bar{D}_{s}}^{2}}\frac{1}{q^{2}-m_{D_{s}}^{2}}\varepsilon_{\mu}(p_{2})F^2(q,\Lambda,m_E), \\  
i\mathcal{M}_b&=&g_{XD_{s0}^*\bar{D}_{s}}g_{D_{s0}^*D_{s}\eta}g_{ \bar{D}_{s}D_{s}^* \eta} \int\frac{d^{4}q}{(2\pi)^{4}}q^{\mu}\frac{1}{k_{1}^{2}-m_{D_{s0}^*}^2}\frac{1}{k_{2}^{2}-m_{\bar{D}_{s}}^{2}}\frac{1}{q^{2}-m_{\eta}^{2}}\varepsilon_{\mu}(p_{2})F^2(q,\Lambda,m_E),  \\  
i\mathcal{M}_c&=&g_{XD_{s0}^*\bar{D}_{s}}g_{D_{s0}^*DK}g_{ \bar{D}_{s} D^* K} \int\frac{d^{4}q}{(2\pi)^{4}}q^{\mu}\frac{1}{k_{1}^{2}-m_{D_{s0}^*}^2}\frac{1}{k_{2}^{2}-m_{\bar{D}_{s}}^{2}}\frac{1}{q^{2}-m_{K}^{2}}\varepsilon_{\mu}(p_{2})F^2(q,\Lambda,m_E),
\end{eqnarray}
\end{widetext}
where   $g$ with subscripts denotes the couplings of the interaction vertices in the triangle diagrams.

 It is well known that many charmonium/charmoniumlike states are produced in $B$ decays, proceeding via the decay $b \to c \bar{c} s$ at the quark level~\cite{Chen:2021erj,Liu:2024uxn}, similar to the production mechanism of the $X(4310)$ in $B$ decays,  as shown in Fig.~\ref{triangleproduction}~(a).     However, the decay $B \to X(4310) K$ poses significant theoretical challenges due to its complex non-perturbative dynamics, making first-principles calculations at the quark level particularly difficult. To account for the significant non-perturbative effects in such heavy hadron weak decays, the FSIs approach has proven to be an essential theoretical tool to understand the hadronization process and final-state rescattering effects that dominate these decays~\cite{Li:1996cj,Du:1998ss,Dai:1999cs,Ablikim:2002ep,Cheng:2004ru,Lu:2005mx,Cao:2023csx}.

 In this letter, we assume that the $B$ meson firstly decays into a  $\bar{D}^*$ meson  and a $D_{s0}^*$ hadronic molecule, and then the $\bar{D}^*$ meson scatters into a $\bar{D}_s$ and a $K$ meson. Finally, the $\bar{D}_sDK$ molecule is dynamically generated by the subsystem $\bar{D}_sD_{s0}^*$ as shown in Fig.~\ref{triangleproduction}(b).  In Ref.~\cite{Faessler:2007cu,Liu:2023cwk}, the production of $D_{s0}^*$ as a $DK$ hadronic molecule in $B$ decays is investigated, which laid the foundation for the present study.  Similarly, the amplitude  of the Feynman diagram of Fig.~\ref{triangleproduction}(b)   is written as
\begin{eqnarray}\label{am3am}
\mathcal{M} &=&  g_{\bar{D}^*\bar{D}_s K} g_{D_{s0}^* \bar{D}_s X}\mathcal{A}({B}\to {D}_{s0}^* \bar{D}^{\ast})^{\mu}p_1^{\nu} \\  \nonumber
 &&\frac{-g_{\mu\nu}+\frac{k_{1\mu}k_{1\nu}}{k_1^2}}{(k_1^2-m_{\bar{D}^*}^2)(k_2^2-m_{D_{s0}^*}^2)(q^2-m_{\bar{D}_s}^2)}F^2(q,\Lambda,m_E).
\end{eqnarray}

The relevant effective Lagrangians and the values of the couplings in Eqs.(\ref{amp1mae}-\ref{am3am}) are listed in the Supplemental Material. 
The partial decay widths of Fig.~\ref{triangle} and Fig.~\ref{triangleproduction}(b)  can be finally obtained as
\begin{eqnarray}
\Gamma=\frac{1}{2J+1}\frac{1}{8\pi}\frac{|\vec{p}|}{{M}^2}\bar{|\mathcal{M}|}^{2},
\end{eqnarray}
where $J$ is the total angular momentum of the initial state $M$, the overline indicates the sum over the polarization vectors of final states, and $|\vec{p}|$ is the momentum of either final state in the rest frame of the $M$ meson. 

In this study, in addition to the uncertainties of parameter $\alpha$ in Eq.~(\ref{formfactor}),   the dominant uncertainties originate from the couplings of the three vertices of the triangle diagrams.  As a result,    we obtain the uncertainties of the
decay widths originating from the uncertainties of these parameters via a Monte Carlo sampling in their 1~$\sigma$ intervals.

In Fig.~\ref{decaywidths}, we show the partial decay widths of $X$ with $M_X=4310$ as a function of $\alpha$, where the uncertainties of the couplings of vertices induce the bands. We can see that the partial widths of the decays $X\to J/\psi \eta$, $X\to \bar{D}_sD_s^*$, and $X\to \bar{D}^*D$     
are of the order of $10^1$, $10^2$, and $10^3$  keV, indicating that the partial decay width of  $X\to \bar{D}^*D$  is larger than those of $X\to \bar{D}_sD_s^*$ and $X\to J/\psi \eta$ by one and two orders of magnitude, respectively.
Taking into the lower and center mass of $X(4310)$, we find that  the ratio of $\Gamma(X\to \bar{D}^*D)/\Gamma(X\to \bar{D}_sD_s^*)$ lies in the range of $9.9 \sim 13.6$, $9.3 \sim 12.6$ , and $8.7\sim11.8$ for the lower, center, and upper mass of $X$, and the corresponding ratio of $\Gamma(X\to \bar{D}_sD_s^*)/\Gamma(X\to J/\psi \eta)$ lies in the range of $5.2\sim 6.1$, $5.1 \sim 5.7$ , and $4.9\sim 5.4$. Regardless of the mass of $X$, the ratio of $\Gamma(X\to \bar{D}^*D):\Gamma(X\to \bar{D}_sD_s^*):\Gamma(X\to J/\psi \eta)\approx50:5:1$.    Therefore, we conclude that $X(4310)$ dominantly decays into $\bar{D}^*D$.

\begin{figure}[!h]
	\centering
	\includegraphics[width=7.5cm]{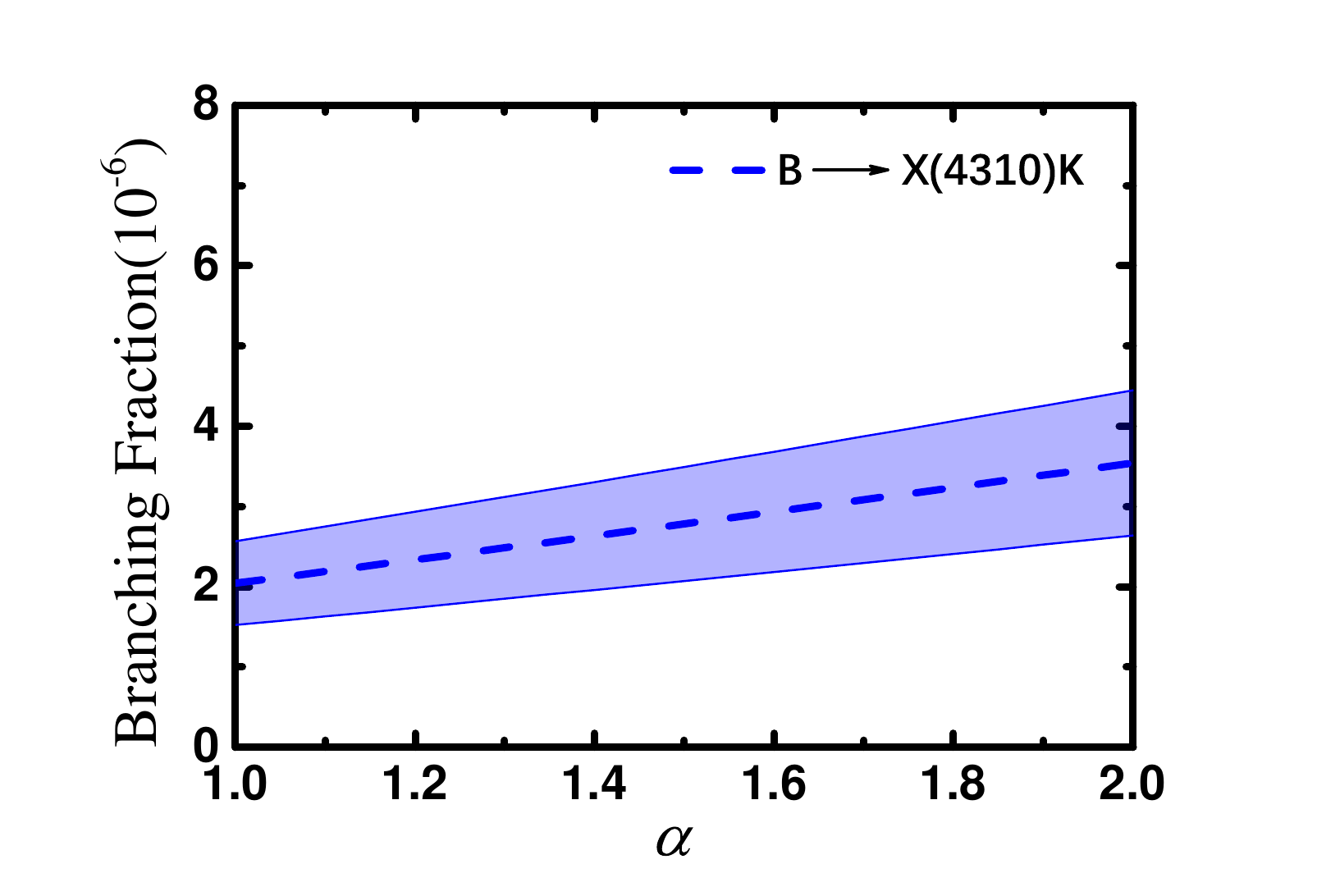}\\
	\caption{Branching fraction of the decays $B \to K X(4310)$  as a function of $\alpha$.  }\label{productionwidths}
\end{figure}

In Fig.~\ref{productionwidths}, we show the branching fraction of the decay $B \to K X$ as a function of $\alpha$, where $m_X=4310$~MeV.  One can see that the branching fraction of the decay $B \to K X(4310)$ is approximately $10^{-6}$. 
Since the production rate of $X(4310)$ in $B$ decays remains of the order of $10^{-6}$ for both its lower and upper mass limits,  the production rate of $X(4310)$  in $B$ decays is of the order of $10^{-6}$. The results of strong decays and production rates of $X(4310)$ for the lower and upper limits  are given in the Supplemental Material.   
Compared with the yield of  $D_{s0}^*(2317)$ in $B$ decays~\cite{Liu:2023cwk},  the production rate of the two-body molecule $DK$ in $B$ decays is larger than that of the three-body molecule $\bar{D}_sDK$ in $B$ decay by three orders of magnitude, consistent with the ratio of the yield of the two-body molecule  $DK$  to that of the three-body molecule $\bar{D}DK$ in $e^+e^-$ collisions~\cite{Wu:2022wgn}, indicating that the ratio of the production rates of the $DK$ molecule to those of the $D\bar{D}_{(s)}K$ molecule is only dependent on the long-range interaction but independent on the short-range interaction in the production process.

Our results indicate that the $X(4310)$ dominantly decays into $\bar{D}^*D$.  In the isospin limit, the branching fraction  of the $X(4310)$ decays into  $\bar{D}^{*0}D^0$, $\bar{D}^0D^{*0}$, $D^{*+}D^-$, and $D^+D^{*-}$ is around $0.25$. As a result, we estimate the branching fractions of the decays $B^+ \to [X(4310) \to {D}^{*\pm}D^\mp ] K^+$ to be  
$5\times 10^{-7}$. Referring to the  branching fractions  $\mathcal{B}(B^+ \to {D}^{*\pm}D^\mp  K^+ )=6\times 10^{-4}$~\cite{BaBar:2010tqo,LHCb:2020qdy}, we estimate the   ratios of $\mathcal{B}[B^+ \to (X(4310) \to {D}^{*\pm}D^\mp ) K^+]/\mathcal{B}(B^+ \to {D}^{*\pm}D^\mp  K^+) \sim 10^{-3}$. The event number of the  decays $B^+ \to {D}^{*\pm}D^\mp  K^+$ of the LHCb Collaboration corresponding to an integrated luminosity of $9$~fb$^{-1}$ is around $2\times 10^{3}$~\cite{LHCb:2024vfz}. One can expect that the event number of the 
 decays $B^+ \to [X(4310) \to {D}^{*\pm}D^\mp ] K^+$  would reach  at least $10$ and  $10^2$  corresponding to  the integrated luminosity of $50$~fb$^{-1}$ and $350$~fb$^{-1}$.   

\begin{figure}[!h]
	\centering
	\includegraphics[width=7.0cm]{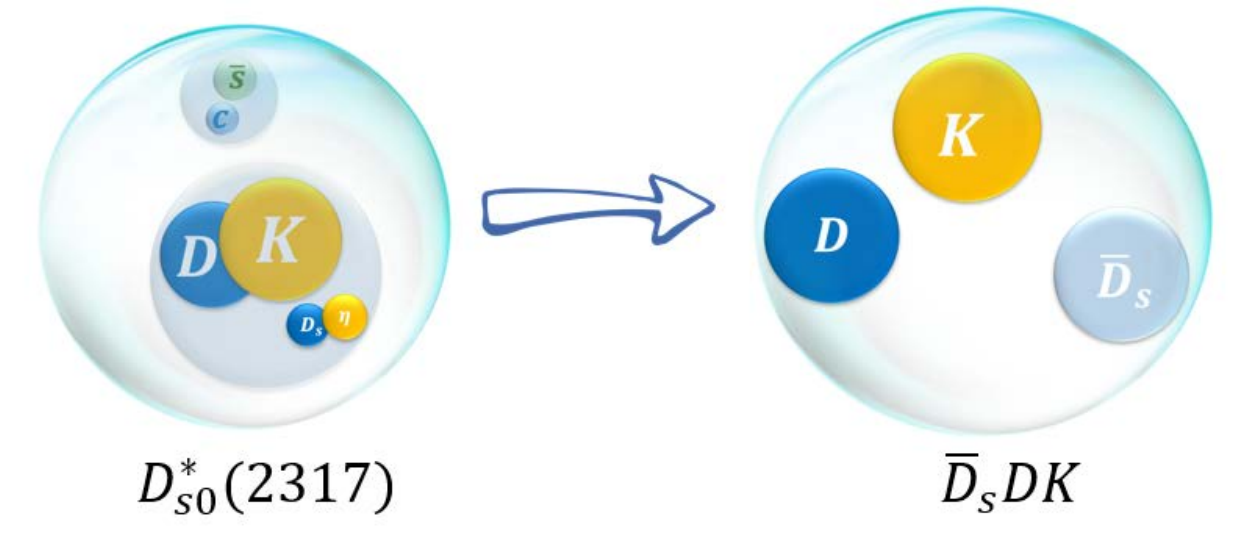}\\
	\caption{ With the $DK$ potential determined by reproducing the mass of the $D_{s0}^*(2317)$  as a mixture of a $DK-D_s\eta$ molecular state and a $c\bar{s}$ bare state, a three-body  $\bar{D}_sDK$ molecule is predicted.   }\label{from2317toX}
\end{figure}

\noindent{\it Summary and outlook.—}
Over the past two decades, the discovery of numerous exotic hadronic states—many considered strong candidates for hadronic molecules—has revolutionized our understanding of hadron-hadron interactions, opening new avenues to explore three-body molecular configurations. In this letter, with the $DK$ interaction determined by the $D_{s0}^*(2317)$ assumed as a mixture state, we predicted a rather special three-body hadronic molecule $J^{PC}=0^{--}$ $\bar{D}_sDK$ with a mass of about $4310$~MeV as shown in Fig.~\ref{from2317toX}. A careful study of its decay behavior and production mechanism shows that it dominantly decays into $\bar{D}^*D$, and its production rate in $B$ decays is of the order of $10^{-6}$. Notably, we found that the ratio of the production rates of the $DK$ molecule to those of the $D\bar{D}_{(s)}K$ molecules is only dependent on the long-range interaction of the production processes. This crucial observation demonstrates that non-perturbative strong interaction effects fundamentally govern the three-body molecular formation.

Moreover,  the event number of the decays $B^+ \to [X(4310) \to {D}^{*\pm}D^\mp ] K^+$ in the LHC corresponding to the integrated luminosity of $50$~fb$^{-1}$ and $350$~fb$^{-1}$ is estimated to be  at least $10$ and $10^2$, respectively.  It should be noted that the LHCb Collaboration observed the signal of $J^{PC}=0^{--}$ charmonium/charmoniumlike states in the decays $B^+ \to  {D}^{*\pm}D^\mp K^+$~\cite{LHCb:2024vfz}.     Therefore, the observation of the $J^{PC}=0^{--}$ $\bar{D}_sDK$ molecule by the LHCb Collaboration in the future is promising.  
On the other hand, the $0^{--}$ $\bar{D}_sDK$ three-body bound state is easily distinguished from charmonium states and two-body molecules. The discovery of the $0^{--}$ $\bar{D}_sDK$ three-body hadronic molecule will extend the configuration of the hadron spectrum and help determine the relevant hadron-hadron interaction and even pinpoint the molecular nature of $D_{s0}^*(2317)$.   We strongly recommend experimental searches for the  $0^{--}$ $\bar{D}_sDK$ molecule in the decay channels   $B^+ \to  {D}^{*\pm}D^\mp  K^+ $.

\noindent {\it Acknowledgement.—}
This work is partly supported by the National Key R\&D Program of China under Grant No. 2023YFA1606703 and the National Natural Science Foundation of China under Grant No. 12435007. Tian-Wei Wu acknowledges support from the National Natural Science Foundation of
China under Grant No.12405108.
  Ming-Zhu Liu acknowledges support from the National Natural Science Foundation of
China under Grant No.12105007.

\bibliography{reference}

\clearpage

\begin{widetext}
\setcounter{page}{1}
\setcounter{figure}{0}  
\setcounter{table}{0}  
\section{Supplemental Material}

We provide details on determining the contact-range potentials and relevant Lagrangians of the Feynman diagrams in Figs.~3 and 4(b) in the main manuscript here. In addition, we present the additional results for the $J^{PC}=0^{-+}$ $\bar{D}_s DK$ system and partial decay widths and production rates of the $X(4310)$ for its lower and upper  mass limits.

\section{ Contact-range EFT Potentials  }

In this letter, we employ the contact-range EFT approach to construct the $DK$, $\bar{D}_sK$, and $D\bar{D}_s$ potentials, which are characterized by an unknown parameter $C_a$~\cite{Hidalgo-Duque:2012rqv,Liu:2023cwk}.  The $DK$ and $\bar{D}_sK$ interactions are determined by reproducing the mass of the $D_{s0}^*(2317)$ and combining SU(3)-flavor symmetry.  The $D\bar{D}_s$ potential is determined by reproducing the mass of $X(3872)$ and combining  SU(3)-flavor symmetry.  Since the $DK$ interaction plays dominant role among these potentials, we study three scenarios to determine  the $DK$ interaction by reproducing the mass of $D_{s0}^*(2317)$, i.e.,  the $D_{s0}^*(2317)$ is assumed as a $DK$ molecule,  a  $D_s\eta-DK$ molecule, and a  mixture of $D_s\eta-DK$ molecular component and bare component.   

The  $DK-D_s\eta$  coupled-channel contact-range  potential  $V$ in matrix form  reads~\cite{Liu:2023cwk}
\begin{equation}
\label{couplepoten}
   V_{DK-D_s\eta}^{J^P=0^+}=\begin{pmatrix} 
   C_a&  -\frac{\sqrt{3}}{2} C_a\\ -\frac{\sqrt{3}}{2}  C_a& 0\end{pmatrix}.
\end{equation}
To estimate  the dressed effect of bare states, in addition to the contact-range potentials in Eq.~(\ref{couplepoten}), we have added an energy-dependent term 
\begin{eqnarray}
V= \beta (s-\bar{s}),    
\end{eqnarray}
where $\bar{s}$ is the energy squared of the mass threshold. The unknown parameter $\beta$ is determined by the weight of  molecular components.  Since recent studies showed that  the molecular component account for more than $70\%$ of the total wave function of $D_{s0}^*(2317)$~\cite{Albaladejo:2016hae,Albaladejo:2018mhb,Ikeno:2023ojl}, we set the weight as $70\%$ to determine the value of $\beta$. 

\subsection{  Momentum Space }

\begin{table}[!h]
 \centering
 \caption{   Values of   the $D_{s0}^*(2317)$ couplings to its constituents ($g$ with subscript), contact range potential ($C_a$), and compositeness of $DK$ and $D_s\eta$ ($P$ with subscript) for a cutoff variation from $0.5$ to $2.0$ GeV. \label{moleculecouplingcutoff} }
\begin{tabular}{c|cccc|ccccc}
 \hline\hline
 Couplings ~~~  &  $\Lambda=0.50$~~~   &    $\Lambda=1.00$~~~ & $\Lambda=1.50$~~~  & $\Lambda=2.00$~~~   &    $\Lambda=0.50$~~~ & $\Lambda=1.00$~~~ &  $\Lambda=1.50$~~~  &  $\Lambda=2.00$~~~  \\
 \hline
 $g_{D_{s0}^*DK}$~(GeV)~~~  & 19.37~~~  &14.72~~~  & 13.32~~~  &12.66~~~& 16.20~~~  & 12.28~~~ & 11.16~~~  & 10.63~~~ \\
 $g_{D_{s0}^*D_s\eta }$~(GeV)~~~  &13.23~~~   &9.54~~~ & 8.40~~~   &7.86~~~&10.42~~~  & 7.70~~~ & 6.89~~~ & 6.50~~~ \\   
  $C_{a}$(fm$^2$)~~~ &-5.78~~~  &-1.84~~~   &-1.03~~~   &-0.71~~~& -6.96~~~  & -2.06~~~ & -1.12~~~ & -0.75~~~ \\ 
 \hline
  Compositeness ~~~ &  $\Lambda=0.50$~~~  &    $\Lambda=1.00$~~~ & $\Lambda=1.50$~~~      &    $\Lambda=1.00$~~~ & $\Lambda=0.50$~~~ & $\Lambda=1.00$~~~ &  $\Lambda=1.50$~~~  &  $\Lambda=2.00$~~~ \\
 \hline
 $P_{DK}$~~~  & 0.92~~~   &0.90~~~ & 0.89~~~  &0.88~~~& 0.65~~~  & 0.63~~~ & 0.62~~~ & 0.62~~~\\
 $P_{D_s\eta }$ ~~~  &0.08~~~ &0.10~~~  & 0.11~~~  &0.12~~~& 0.05~~~  & ~0.07~~~ & 0.08~~~ & 0.08~~~  \\
\hline\hline
 \end{tabular}
 \end{table}

Assuming the $D_{s0}^*(2317)$ as a $DK$ bound state, one can determine the   value  of $C_{a}=-0.98$~fm$^2$ for a cutoff $\Lambda=2.0$ GeV,  $C_{a}=-1.41$~fm$^2$ for a cutoff $\Lambda=1.5$ GeV,  $C_{a}=-2.44$~fm$^2$ for a cutoff $\Lambda=1.0$ GeV,  and $C_{a}=-7.21$~fm$^2$ for a cutoff $\Lambda=0.5$ GeV. Then, assuming that the $D_{s0}^*(2317)$ is dynamically generated by $DK$ and $D_s \eta$ coupled channels, one can determine the  absolute value  of $C_a$ as a function of cutoff in the left panel of  Table~\ref{moleculecouplingcutoff}. One can see that the size  (absolute value) of $C_a$ decreases, and the $DK$ and $D_s\eta$ components account for around $90\%$ and $10\%$ of the total wave function, respectively. Finally, when taking into account the effect of bare states, we find that the size (absolute value) of $C_a$ increases, a bit smaller than the first scenario. In the third scenario,   the $DK$ and $D_s\eta$ components account for around $63\%$ and $7\%$ of the total wave function.   After the $DK$ interaction is determined,  the $\bar{D}_sK$ potential is determined as   half of the $DK$ potential from the SU(3)-flavor symmetry, i.e., $V_{\bar{D}_sK}=\frac{1}{2}V_{DK}$~\cite{Guo:2006fu,Guo:2009ct,Liu:2012zya,Altenbuchinger:2013vwa}.  


For the scattering process $\bar{D}_s D\rightarrow \bar{D}_sD $, the contact potential is characterized by the parameter $C_{1a}$~\cite{Hidalgo-Duque:2012rqv}, which is the same as the isovector contact potential of  $\bar{D} D\rightarrow \bar{D}D $ under SU(3)-flavor symmetry. In the following, we analyze the relationship of the parameters of the contact-range EFT by the light meson saturation approach~\cite{Peng:2020xrf}. The contact-range potentials of the isoscalar $\bar{D}^{(*)}D^{(*)}$ system are parameterized as $C_{0a}$ and $C_{0b}$, and the isosvector $\bar{D}^{(*)}D^{(*)}$ system as $C_{1a}$ and $C_{1b}$. According to the light meson saturation approach, we have the ratios $C_{0a}:C_{0b}:C_{1a}:C_{1b}=1:0.35:0.42:0$. Identifying $X(3872)$ as the bound state of $J^{PC}=1^{++}\bar{D}^*D$, one can obtain the sum of $C_{0a}+C_{0b}=-0.79$~fm$^2$ for a cutoff $\Lambda=1$~GeV, then further obtain the values of $C_{0a}=-0.58$~fm$^2$, $C_{0b}=-0.21$~fm$^2$, and $C_{1a}=-0.24$~fm$^2$. Similarly, we obtain the value of $C_{1a}=-0.60$~fm$^2$ for a cutoff $\Lambda=0.5$~GeV, $C_{1a}=-0.15$~fm$^2$ for a cutoff $\Lambda=1.5$~GeV, and $C_{1a}=-0.11$~fm$^2$ for a cutoff $\Lambda=2.0$~GeV.              
In Ref.~\cite{Ji:2022vdj}, Ji et al. obtained the value of $C_{1a}=-0.33\pm0.02$~fm$^2$ for a cutoff $\Lambda=1$~GeV by simulating the mass distributions of $\bar{D}_{(s)} D_{(s)} $ of the processes of $\gamma\gamma \to \bar{D}D$ and $B^+ \to K^+ {D}_{(s)}^+{D}_{(s)}^-$, consistent with the analysis of the light meson saturation approach.   A recent approach analyzing the Lattice QCD data of $\bar{D} D-\bar{D}_s D_s $ coupled-channel scattering, the value of $C_{1a}$ is estimated to be in the range of  $-0.44\sim-0.64$~fm$^2$~\cite{Shi:2024llv},  it's absolute value a bit larger than our estimations and Ref.~\cite{Ji:2022vdj}. The ratio of the $DK$ potential in the first scenario to the $\bar{D}_sD$ potential is from $8.9$ to $12.0$, and the ratio of the $DK$ potential in the third scenario to the $\bar{D}_sD$ potential is from $6.8$ to $11.6$. The average of the above ratios is around $10$.  
In the momentum space, we establish the following approximate ratio for the contact potentials at the cutoff $\Lambda=1$~GeV,  i.e.,  $C_a^{DK}:C_a^{\bar{D}_s K} : C_a^{\bar{D}_s D}=1:0.5:0.1$.

\subsection{ Coordinate Space}

\begin{table}[!h]
 \centering
 \caption{Masses of $DK-D_s\eta$ and $DK$ molecules  (in units of MeV), and probabilities in $D_{s0}^*(2317)$ based on the different molecular components of $D_{s0}^*(2317)$.\label{twobody} }
 \begin{tabular}{c|c ccc cc}
 \hline\hline
 Components of $D_{s0}^*(2317)$&$M(DK-D_s\eta$) &    $M(DK)$ & $M(c\bar{s}$) \cite{Yang:2021tvc} &  $P(c\bar{s})$~~~  &   $ P(DK)$&$ P(D_s\eta)$\\
 \hline
70\% molecule+30\% $c\bar{s}$ &$2280$& 2349  &2406& 30\%& 60\% & 10\%  \\
 \hline
100\% molecule&2318& 2358  &2406& 0\%& 90\% & 10\%  \\
 \hline
 {50\% molecule+50\% $c\bar{s}$}&
   2230& 2336 &2406& 50\%& 42\% & 8\% \\
\hline\hline
 \end{tabular}
 \end{table}

The above contact-range potentials are transformed into $C_a\delta^3(\vec{r})$ in the coordinate space~\cite{Liu:2018zzu}. 
The strength $C_a$ of the $DK$ interaction can be determined by the 
leading-order chiral perturbation theory~\cite{Wu:2019vsy}, i.e., the Weinberg-Tomozawa term $C_a^{DK}=-\frac{C_W}{2f_{\pi}^2}$, where   $C_W=2$ and $f_{\pi}=130$ MeV~\cite{Liu:2012zya},    resulting in the value of $C_a^{DK}=-2.24$ $\rm{fm}^2$.  Assuming the $D_{s0}^*(2317)$ as a mixture of a $DK-D_s{\eta}$ molecule and a $c\bar{s}$ bare state, $R_c$ is determined as $0.472$ fm.  {As indicated in Ref.~\cite{Liu:2018zzu}, the result of $R_c=0.5$ fm in coordinate space is equal to the result of $\Lambda=1$~GeV in momentum space. The $DK$ potential determined for $R_c=0.472$ fm should be consistent with that for $\Lambda=1$~GeV. As shown in Table~\ref{moleculecouplingcutoff}, the $DK$ potential in the third scenario is determined as $C_a^{DK}=-2.06$ $\rm{fm}^2$, consistent with the value derived in coordinate space.}        
With such a $DK$ potential, we found a bound state with a binding energy of $14$ MeV, which indicates that the obtained $DK$ potential is less attractive than that of assuming $D_{s0}^*$ as a pure $DK$ bound state, consistent with the conclusion derived in momentum space.     
In addition, we vary the molecule compositeness of $D_{s0}^*(2317)$ for which $R_c$ ranges from $ 0.438$ to $0.513$ from $50\%$ to $100\%$ to estimate the uncertainties of the input. Accordingly, we can obtain the corresponding $DK$ interaction, of which $R_c$ ranges from $0.438-0.513$ fm. The 
results are summarized in Table~\ref{twobody}.  In a recent measurement by the BESIII Collaboration, the branching fraction $\mathcal{B}(D_{s0}^*(2317) \to \pi^0 D_s)$ was first determined to be approximately $1$~\cite{BESIII:2017vdm}, which implies that the branching fraction $\mathcal{B}(D_{s0}^*(2317) \to D_s \gamma)$ is minor. The ratio of  $\mathcal{B}(D_{s0}^*(2317) \to D_s \gamma)/\mathcal{B}(D_{s0}^*(2317) \to \pi^0 D_s)$ is estimated to be around $10^{-2}$ with for a molecular state~\cite{Faessler:2007gv,Fu:2021wde}  and $0.15$ at least for an $c\bar{s}$ bare state~\cite{Godfrey:2003kg,Colangelo:2003vg,Colangelo:2005hv}.    
The measurement of the branching fraction $\mathcal{B}(D_{s0}^*(2317) \to \pi^0 D_s)\sim 1$  by  the BESIII Collaboration  supports the molecular interpretation of $D_{s0}^*(2317)$, which favors  our hypothesis that the $D_{s0}^*(2317)$  represents an  admixture of a molecular component and a bare state, with the molecular configuration constituting the dominant fraction.      Once the $DK$ potential is determined,   the $\bar{D}_sK$ potential is determined as  half of the $DK$ potential from the SU(3)-flavor symmetry, i.e., $V_{\bar{D}_sK}(r)=\frac{1}{2}V_{DK}(r)$~\cite{Guo:2006fu,Guo:2009ct,Liu:2012zya,Altenbuchinger:2013vwa}.

By fitting $X(3872)$ with the contact-range EFT, the strength of the $\bar{D}^*D$ potential $C_a^{\bar{D}^*D}$ is about $-0.79$ $\rm{fm}^2$ and $R_c=0.434$ fm. According to the light meson saturation approach~\cite{Peng:2020xrf}, we have $C_a^{\bar{D}^*D}:C_a^{\bar{D}_sD}=1.35:0.42$, resulting $C_a=-0.24$ $\rm{fm}^2$. In the coordinate space, we establish the following approximate ratio for the contact potentials at the cutoff $\Lambda=1$~GeV,  i.e.,  $C_a^{DK}:C_a^{\bar{D}_s K} : C_a^{\bar{D}_s D}=1:0.5:0.1$.

\section{Effective Lagrangians}

The Lagrangians describing the interactions between  the hadronic molecules   and their constituents are written as 
\begin{eqnarray}
\mathcal{L}_{XD_{s0}^*\bar{D}_{s}}&=& g_{XD_{s0}^*\bar{D}_{s}}XD_{s0}^*\bar{D}_{s},     \\ \nonumber
\mathcal{L}_{D_{s0}^*D_{s}\eta}&=& g_{D_{s0}^*D_{s}\eta}D_{s0}^*D_{s}\eta,
  \\ \nonumber
\mathcal{L}_{D_{s0}^*DK}&=& g_{D_{s0}^*DK}D_{s0}^*DK,
\end{eqnarray}
where the molecule couplings to their constituents are determined by  the residues of the pole obtained by solving the Lippmann-Schwinger  
equation~\cite{Liu:2023cwk}.  In this letter, we take the mass of  $X$  under the $DK\bar{D}_s$ mass threshold  $22^{+24}_{-14}$ MeV, where the $D_{s0}^*\bar{D}_s$ sub-system account for around $80\%$ of the total wave function.    Therefore, the couplings of $g_{XD_{s0}^*\bar{D}_{s}}$, $g_{D_{s0}^*D_{s}\eta}$, and $g_{D_{s0}^*DK}$ are estimated by the contact-range EFT as shown  in Table~\ref{moleculecouplingXX}, where the uncertainties of the couplings come from the cutoff varying from $1$ to $2$ GeV.     
It should be noted that the $D_{s0}^*$ is not the $D_{s0}^*(2317)$ instead of  a  $DK$ bound state with a binding energy of $22^{+24}_{-14}$ MeV.  

\begin{table}[!h]
 \centering
 \caption{ Values of the hadronic molecules couplings to their constituents (in units of GeV).  \label{moleculecouplingXX} }
 \begin{tabular}{c|ccccccc}
 \hline\hline
 Couplings ~~~  &    $X=4.286$~~~ & $X=4.310$~~~ &  $X=4.324$~~~    \\
 \hline
  $g_{XD_{s0}^*\bar{D}_{s}}$~~~   &$21.72\pm 4.28$~~~ &$15.86\pm2.34$~~~  & $9.86\pm0.81$~~~    \\ 
 $g_{D_{s0}^*DK}$~~~   &$14.61\pm2.03$~~~& $11.39\pm1.17$~~~  & $7.79\pm1.08$~~~  \\
 $g_{D_{s0}^*D_s\eta }$~~~   &$9.47\pm1.65$~~~ &$7.51\pm1.11$~~~  & $4.93\pm0.27$~~~  \\   
\hline\hline
 \end{tabular}
 \end{table}

The Lagrangian describing the charmed mesons couplings to one light meson and $J/\psi$ are written as~\cite{Oh:2000qr}
\begin{eqnarray}
\mathcal{L}_{\psi \bar{D}_{s} D_{s}}&=& i g_{\psi \bar{D}_{s}D_{s}} \psi_{\mu} (\partial^{\mu} D_{s} \bar{D}_{s}^{\dag}- D_{s} \partial^{\mu}\bar{D}_{s}^{\dag}), \\ \nonumber  
\mathcal{L}_{D_s D_s^{\ast} \eta}&=& -i g_{D_s D_s^{\ast} \eta} (D_s \partial^{\mu}\eta D^{\ast\dag}_{s\mu}-D_{s\mu}^* \partial^{\mu} \eta  D_s^{\dag}),  \\ \nonumber  
\mathcal{L}_{D_s D^{\ast} K }&=& -i g_{D_s D^{\ast} K} (D_s \partial^{\mu}K  D^{\ast\dag}_{\mu}-D_{\mu}^* \partial^{\mu} K  D_s^{\dag}),
\end{eqnarray}
where the couplings are  determined as $g_{\psi \bar{D}_{s}D_{s}}=5.8\pm 0.9$, $g_{ {D}_{s}D_{s}^* \eta}=5.72\pm0.58$, and $g_{ {D}_{s}D^* K}=14.00\pm1.73$  via SU(3)-flavor symmetry~\cite{Bracco:2011pg}. 

The amplitude of the  weak decay $B(k_0) \to D_{s0}^*(q_1) \bar{D}^*(q_2)$, described by the naive factorization approach~\cite{Liu:2023cwk},
is written as
\begin{align}\label{am3}
\mathcal{A}({B}\to {D}_{s0}^* \bar{D}^{\ast}) =& \frac{G_{F}}{\sqrt{2}}V_{cb}V_{cs} a_{1} f_{{D}_{s0}^* }\{-q_{1}\cdot \varepsilon(q_{2})  (m_{ {D}^{\ast}}+m_{{B}})A_{1}\left(q_{1}^{2}\right) +(k_{0}+q_{2}) \cdot \varepsilon(q_{2}) q_{1}\cdot (k_{0}+q_{2}) \\ \nonumber       & 
\frac{A_{2}\left(q_{1}^{2}\right)}{m_{ {D}^{\ast}}+m_{{B}}} +(k_{0}+q_{2}) \cdot \varepsilon(q_{2}) [(m_{{D}^{\ast}}+m_{{B}})A_{1}(q_{1}^2)  -(m_{{B}}-m_{ {D}^{\ast}})A_{2}(q_1^2) -2m_{{D}^{\ast}} A_{0}(q_{1}^2)]  \} ,
\end{align}
 where $G_F = 1.166 \times 10^{-5}~{\rm GeV}^{-2}$, $V_{cb}=0.0395$, $V_{cs}=0.991$, $f_{D_{s0}^*}=59$~MeV, and  $a_1=1.07$.   The form factors of $A_{0}(t)$, $A_{1}(t)$, and $A_{2}(t)$ with $t \equiv q_1^{2}$ can be parameterized as~\cite{Verma:2011yw} 
\begin{equation}
X(t)=\frac{X(0)}{1-a\left(t / m_{B}^{2}\right)+b\left(t^{2} / m_{ B}^{4}\right)}.
\end{equation}
For these form factors, we adopt those of the covariant light-front quark model, i.e.,
$(A_0(0), a, b)^{B \to \bar{D}^{\ast}} = (0.68, 1.21, 0.36)$, $(A_1(0), a, b)^{B \to \bar{D}^{\ast}} = (0.65, 0.60, 0.00)$,  and  $(A_2(0), a, b)^{B \to \bar{D}^{\ast}} = ( 0.61, 1.12, 0.31)$~\cite{Verma:2011yw}. Following Ref.~\cite{Wu:2023rrp}, the $B \to \bar{D}^*$ transition results in a $10\%$ uncertainty.

\section{ Additional results   }

\subsection{  Results of $J^{PC}=0^{-+}$ $\bar{D}_sDK$ molecule }

\begin{table}[!h]
 \centering
 \caption{  Binding energy (in MeV), weights of Jacobian channels, root mean square radii (in fm) and the expectation values of the Hamiltonian (in MeV)  of $0^{-+}$ $\bar{D}_sDK$.\label{threebody0-+} }
 \begin{tabular}{c|  ccc c}
 \hline\hline
 Sets&  B.E.($0^{-+})$ &    $ P_{\bar{D}_sK-D}$& $P_{DK-\bar{D}_s}$ & $ P_{\bar{D}_sD-K}$\\
 \hline
$\alpha=1$ & $26^{+22}_{-16}$& $13^{+0}_{-1}$ \%&$76^{-0}_{+0}$ \%&$11^{-0}_{+1}$ \%\\
 $\alpha=2$&  $28^{+23}_{-17}$& $14^{+0}_{-1} \%$& $74^{+1}_{+1}$ \%&$12^{-1}_{+0}$ \%\\
 \hline
  Scenarios&  $r_{\bar{D}_sK}$ &  $r_{DK}$  & $r_{\bar{D}_sD}$ &$\langle T \rangle$\\
 \hline
$\alpha=1$ & $1.5_{+0.7}^{-0.3}$  &$1.1_{+0.5}^{-0.2} $  &$1.3_{+0.6}^{-0.3}$&$187_{-73}^{+78} $  \\
 $\alpha=2$& $1.4_{+0.6}^{-0.3}$  & $1.1_{+0.4}^{-0.2} $  &$1.2_{+0.5}^{-0.2}$&$195_{-73}^{+77}$ \\
 \hline
  Scenarios& $ \langle V_{D_s\bar{K}}\rangle$~~~ &  $\langle V_{DK}\rangle~~~$  &   $ \langle V_{D_s\bar{D}}\rangle$&$ \langle V^{C=+}_{D_s\bar{D}^*_{s0}}\rangle$\\
 \hline
$\alpha=1$ &$-44_{+20}^{-24}$& $-151_{+61}^{-71}$&$-16_{+6}^{-6}$&$-2_{+1}^{-0}$\\
 $\alpha=2$&  $-47_{+21}^{-24}$& $-154_{+61}^{-70}$&$-17_{+6}^{-6}$&$-4_{+2}^{-1}$\\
 \hline
 \hline
 \end{tabular} 
\end{table}

As shown in Table~\ref{threebody0-+}, the $0^{-+}$ state is a bit more bound than the $0^{--}$ state due to the attractive interaction in positive $C$-parity, which is about  $27^{+24}_{-17}$ MeV. The weights of Jacobi channels $i=1-3$ are about 13\%, 75\% and 12\% of $D_s\bar{K}-\bar{D}$, $\bar{D}\bar{K}-D_s$ and $D_s\bar{D}-\bar{K}$ respectively. The root mean square radii (in units of fm) and the expectation values of the Hamiltonian (in units of MeV) of the bound state $0^{-+}$ $\bar{D}_sDK$ are also shown in Table~\ref{threebody0-+}.

\subsection{  Results of strong decays and production rates }

\begin{figure}[!h]
	\centering
	\includegraphics[width=15cm]{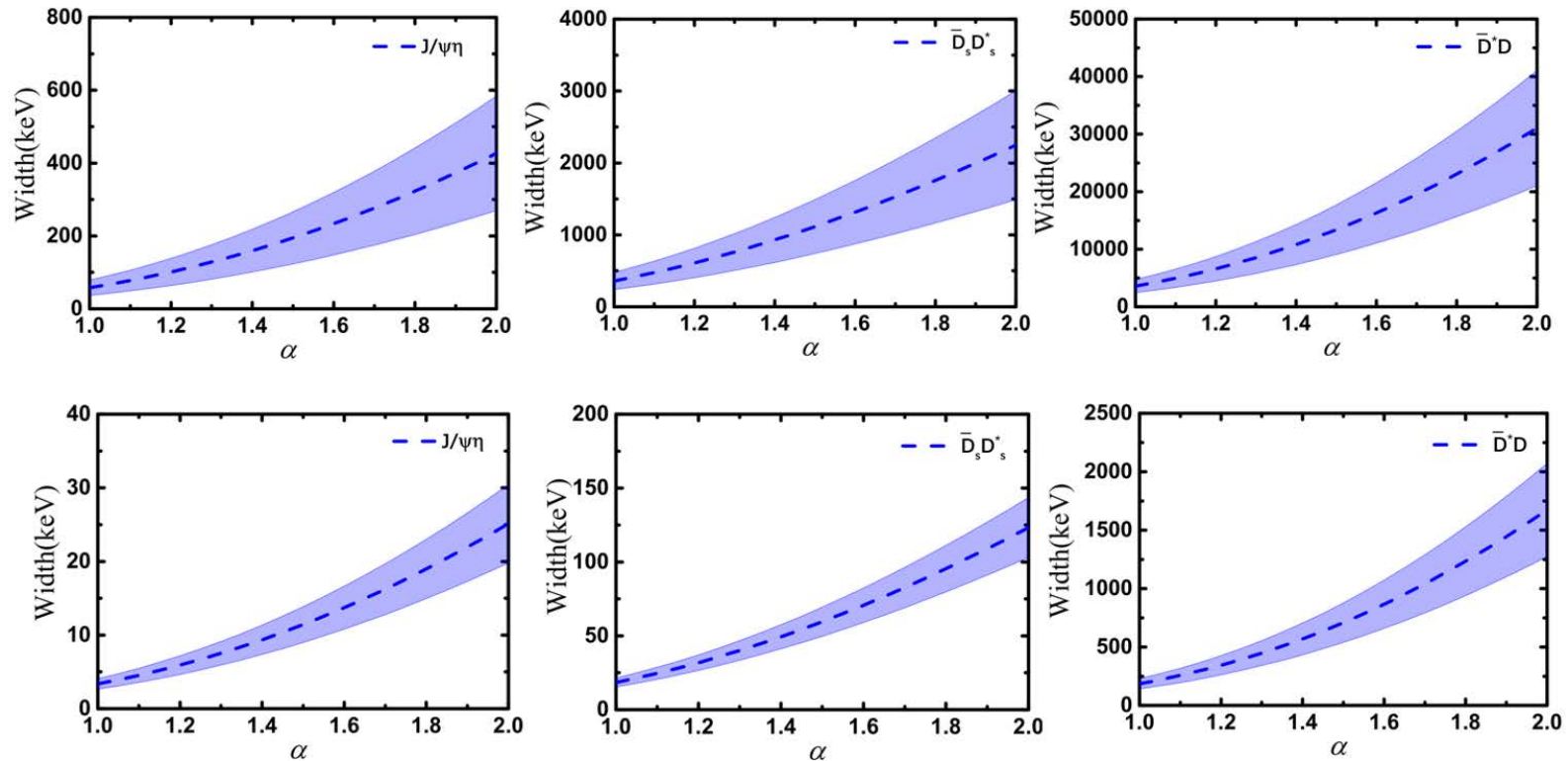}\\
	\caption{Partial widths of the decays $X\to J/\psi \eta$, $X\to \bar{D}_sD_s^*$, and $X\to \bar{D}^*D$ as a function of $\alpha$. The upper and lower panels correspond to the mass of $X=4286$~MeV and  $X=4324$~MeV.    }\label{updecaywidths}
\end{figure}

The main text presented only the partial decay widths and production rates for the mass  $X=4310$ MeV. Here, we show  the partial widths of the decays $X\to J/\psi \eta$, $X\to \bar{D}_sD_s^*$, and $X\to \bar{D}^*D$ as  functions of $\alpha$ for  $X=4286$~MeV and  $X=4324$~MeV(upper and lower panels of Fig.~\ref{updecaywidths}). Although the absolute values of the partial widths vary significantly, their ratios remain relatively stable.   Similarly, we present the production rates of $X$ in $B$ decays for $X=4286$~MeV and $X=4324$~MeV(left and right panels of Fig.~\ref{upproductionrates}). The production rates are of the order of $10^{-6}$.             

\begin{figure}[!h]
	\centering
	\includegraphics[width=12cm]{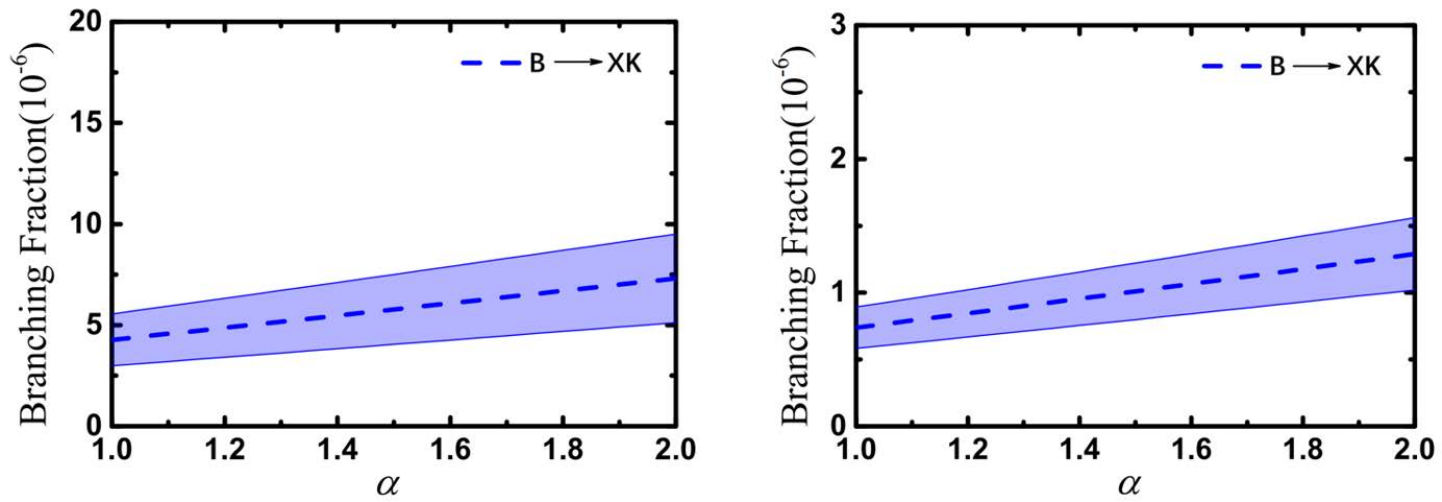}\\
	\caption{ Branching fraction of the decays $B \to K X$  as a function of $\alpha$. The left and right panels correspond to the results for the mass of $X=4286$~MeV and  $X=4324$~MeV.     }\label{upproductionrates}
\end{figure}

Our results indicate that the branching fraction of  $X(4310) \to J/\psi \eta$ can reach  up to the order of $10^{-2}$, yielding the branching fraction $\mathcal{B}[B^+ \to (X(4310) \to J/\psi \eta) K^+]\sim 10^{-8}$. Given the measured branching fraction $\mathcal{B}(B^+ \to J/\psi \eta  K^+)=10^{-4}$~\cite{BaBar:2004iez,Belle:2013vio},    
the ratio  $\mathcal{B}[B^+ \to (X(4310) \to  J/\psi \eta ) K^+]/\mathcal{B}(B^+ \to J/\psi \eta  K^+)$ is  approximately $10^{-4}$.  The LHCb Collaboration observed around $5\times 10^3$ events for   $B^+ \to J/\psi \eta  K^+$ with an integrated luminosity of $9$~fb$^{-1}$~\cite{LHCb:2022oqs}. Extrapolating to $350$~fb$^{-1}$, the expected event count would reach about   $10$  for $B^+ \to [X(4310) \to  J/\psi \eta ] K^+$. For the decay $B^+ \to [X(4310) \to  D_s^{*+}D_{s}^-] K^+ $, the dominant decay  $D_s^{*+} \to D_s^{+}\gamma$ significantly reduces detection efficiency of $D_s^{*+}$ meson at LHCb. Furthermore, the experimental branching fraction for $B^+ \to   D_s^{*+}D_{s}^- K^+ $ is still missing~\cite{ParticleDataGroup:2024cfk}.     Thus, the most promising channel to observe the $J^{PC}=0^{--}$ $\bar{D}_sDK$ molecule is $B^+ \to [X(4310)\to{D}^{*\pm}D^\mp] K^+$.

\end{widetext}

\end{document}